\documentstyle[12pt,preprint,revtex]{aps}
\begin{document}
\def\overlay#1#2{\setbox0=\hbox{#1}\setbox1=\hbox to \wd0{\hss #2\hss}#1%
\hskip -2\wd0\copy1}
\rightline{To be published in JETP {\bf 104},\ No.4(10) (1993)}
\begin{title}
\begin {center}
{ON THE THEORY OF ``ODD''-PAIRING SUPERCONDUCTORS.}
\end{center}
\end{title}
\author{E.Z.Kuchinskii,\ M.V.Sadovskii,\ M.A.Erkabaev}
\begin{instit}
Institute for Electrophysics,\ Russian Academy of Sciences,\\
Ural Branch,\ Ekaterinburg 620219,\ Russia \\
E-mail: sadovski@ief.e-burg.su
\end{instit}
\begin{abstract}
We consider the model of superconducting pairing with the energy gap function
which is odd over $k-k_{F}$.\ In this case superconductivity is possible
even in the presence of an arbitrarily large point-like repulsion between
electrons,\ which is attractive for the theory of high-temperature
superconductors.\ We discuss mainly a model pairing interaction for which the
BCS equations can be solved exactly,\ allowing the complete analysis of the
interplay of the usual (``even'') and ``odd'' pairing,\ depending on the
coupling constant.\ We also show that the normal impurities (disorder) lead
to a rapid degradation of the ``odd'' pairing,\ which is even more rapid than
in case of magnetic impurities in traditional superconductors.
\end{abstract}

\newpage
\vskip 0.5cm
\narrowtext
\section{INTRODUCTION}

Traditional BCS theory of superconductivity\cite{Genn} is based upon an
assumption of the existence of an attractive interaction of electrons with
opposite spins close to the Fermi level. Usually it is assumed that this
attractive interaction overcomes the Coulomb repulsion of electrons at least
in some part of the phase space,\ which is considered to be the necessary
condition for superconducting state at low temperatures.\  Naturally,\ the
strong repulsion of electrons characteristic to Hubbard-like models,\ which
are extensively used to describe the electronic properties of metallic oxides,\
is a factor opposing superconductivity.\ From this point of view it is
interesting to consider models,\ in which the influence of such repulsion is
strongly suppressed or even absent.\ A model of this kind was proposed in a
recent paper by Mila and Abrahams\cite{MA}. It is based on the assumption that
the BCS gap  $\Delta(\vec k,\omega)$ depends only on $|\vec k|$,\ or more
precisely on the quasi-particle energy $\xi_{k}=v_{F}(|\vec k|-k_{F})$,\
measured from the Fermi level and is an {\em odd} function of this variable.\
In this case superconductivity becomes possible even for the arbitrarily
strong point-like repulsion between electrons.\ Physically it is obvious that
such a state may realize in case of strong enough repulsion,\ when the usual
(``even'') superconductivity is suppressed.\  References to the earlier
works on the ``odd''-pairing can be found in Ref.\cite{Abr}.

The aim of our paper is more detailed than in Ref.\cite{MA} study of the
``odd'' pairing and its relation to the usual ``even'' case,\ using the
simplest weak coupling approximation and a model pairing interaction which
allows the exact solution of BCS equations.\ This model allows the detailed
analytic analysis and comparison with the results of Ref.\cite{MA} which were
obtained numerically for more ``realistic'' interaction.\ Besides that we
consider the influence of normal impurities on the ``odd'' pairing.\ This
influence is unusually strong \cite{KS} and superconductivity is completely
suppressed even faster than in the case of magnetic impurities in traditional
superconductors.\ We solve this problem both using the model pairing
interaction and numerically in the case of ``realistic'' interaction proposed
in Ref.\cite{MA}.\ Despite the obvious attractiveness of the odd pairing for
the analysis of high-temperature superconductivity in oxides its strong
instability towards disordering makes it unlikely candidate for pairing in
cuprates.

\section{EQUATIONS FOR THE GAP AND \\ TRANSITION TEMPERATURE}

The model is based on the demonstration of the fact \cite{MA} that the weak
coupling BCS gap equation\cite{Genn}:
\begin{equation}
\Delta(\xi)=-N(0)\int\limits_{-\infty}^{\infty}d{\xi'}V(\xi,\xi')
\frac{\Delta(\xi')}{2\sqrt{\xi'^{2}+\Delta^{2}(\xi')}}
th\left(\frac{\sqrt{\xi'^{2}+ \Delta^{2}(\xi')}}{2T}\right) \label{1}
\end{equation}
can acquire the nontrivial solution $\Delta(\xi)=-\Delta(-\xi)$
(i.e.\ odd over $k-k_{F}$,\ $\xi=v_{F}(k-k_{F})$) in case of the
presence in $V(\xi,\xi')$ of a strong enough attractive part even if
very strong (infinite) point-like repulsion is also present.\ It is easily seen
\cite{MA} that for the odd $\Delta(\xi)$ the repulsive part of interaction in
(\ref{1}) just drops out while the attractive part $V_{2}(\xi,\xi')$
may lead to the pairing with some unusual properties:\
the gap $\Delta(\xi)$ is zero at the Fermi surface which leads to the
gapless superconductivity.\ We must stress that in this case we have isotropic
pairing and the gap is zero everywhere at the Fermi surface which is different
from the case of anisotropic,\ e.g.\ $d$-wave pairing\cite{Abr}.

Thus,\ in the following we assume that the interaction kernel in Eq.(\ref{1})
consists of two parts:\
$V(\xi ,\xi ')=V_{1}(\xi ,\xi ')+V_{2}(\xi ,\xi ')$,\ where
\begin{equation}
V_{1}(\xi ,\xi ')=\left\{
\begin{array}{lcl}
U>0&\mbox{ for }&|\xi|,|\xi '|<E_{F} \\
0&\mbox{ for }&|\xi|,|\xi '|>E_{F}
\end{array}
\right.
\label{V1}
\end{equation}
---is the point-like repulsion of electrons,\ while $V_{2}(\xi ,\xi ')$
---is an effective pairing interaction (attraction),\ which is nonzero for
$|\xi|,|\xi '|<\omega_{c}$ and $|\xi -\xi '|<\omega_{c}$
(the last condition is of the major importance)\, where $\omega_{c}\ll E_{F}$
---plays the role of characteristic frequency of Bosons,\ responsible for the
pairing.\ The pairing ``potential'' $V_{2}(\xi ,\xi ')$ can be modelled by
various functions,\ e.g. by a step-function as in Ref.\cite{MA}.
In Ref.\cite{MA} the major attention was to the following form of this
interaction:
\begin{equation} V_{2}(\xi ,\xi
')=\left\{ \begin{array}{lcl} -V(|\xi -\xi '|/\omega_{c})^{-2/3}&\mbox{ for
}&|\xi|,|\xi '|,|\xi -\xi '|<E_{F} \\ 0&\mbox{ for }&|\xi|,|\xi '|\mbox{ or
}|\xi -\xi '|>E_{F} \end{array} \right.  \label{V2-2/3}
\end{equation}
which was justified by the wish to get the following tunneling density of
states of a superconductor:\  $N_{T}(E)\sim |E|^{2}$ for $E\rightarrow 0$.
The integral gap equation (\ref{1}) with such $V_{2}(\xi ,\xi ')$ was solved
in Ref.\cite{MA} numerically,\ and some other results qualitatively similar to
those observed in high-temperature superconducting oxides were obtained.

Here we shall concentrate mainly on the following model interaction\cite{KS}:
\begin{equation}
V_{2}(\xi ,\xi ')=\left\{ \begin{array}{lcl}
-Vcos\left(\frac{\pi}{2}\frac{\xi -\xi '}{\omega_{c}}\right)&\mbox{ for
}&|\xi|,|\xi '|,|\xi -\xi '|<\omega_{c} \\ 0&\mbox{ for }&|\xi|,|\xi '|\mbox{
or}|\xi -\xi '|>\omega_{c}
\end{array} \right.  \label{V2-cos}
\end{equation}
The major advantage of such a choice is that now we can reduce the integral
gap equation to a transcendental one and solve it easily.\ The choice of Eq.
(\ref{V2-cos}) is not in this sense unique,\ quite a number of similar
``potentials'' may be proposed.\ We can use for example
$V_{2}(\xi ,\xi ')\sim ch(\xi -\xi ')$ or $(\xi -\xi ')^{2}$.
However,\ in case of Eq.(\ref{V2-cos}) we obtain the results which in some
sense are closest to those obtained with the ``realistic'' choice of
(\ref{V2-2/3}) (e.g.\ for the density of states).\ The majority of qualitative
conclusions below do not depend at all on the form of the model ``potential''.
The importance of the model interaction of Eq.(\ref{V2-cos}) is obviously
related to the fact that any ``potential'' which is an even function of
$\xi-\xi'$ on the interval from $-\omega_{c}$ to $\omega_{c}$,\ can be
represented by the Fourier expansion over cosines.\ In this sense our analysis
forms the basis for rather general study.

We should like to note,\ that the real choice of the pairing interaction
$V(\xi ,\xi ')$ can be made with the use of dielectric function formalism of
superconductivity theory\cite{KMK,PVTSP},\ where rather general expressions
for the interaction kernel of the BCS theory are obtained.\ Unfortunately,\
it is unclear how we can get any nontrivial dependence of this kernel upon
$|\xi -\xi '|$,\ because in this formalism we typically get the dependence of
$V(\xi ,\xi ')$ on $|\xi |$ and $|\xi '|$ separately\cite{KMK,PVTSP}.

Superconducting transition temperature is determined by the linearized
equation:
\begin{equation}
\Delta(\xi)=-N(0)\int\limits_{-\infty}^{\infty}d{\xi '} V(\xi ,\xi ')
\frac{\Delta(\xi ')}{2\xi '} th\left(\frac{\xi '}{2T_{c}}\right)
\label{5}
\end{equation}
Using (\ref{V1}) and (\ref{V2-cos}) we get:
\begin{equation}
\Delta(\xi)=g\int\limits_{-\omega_{c}}^{\omega_{c}}d{\xi '}
cos\left(\frac{\pi}{2}\frac{\xi -\xi '}{\omega_{c}}\right)
\frac{\Delta(\xi ')}{2\xi '} th\left(\frac{\xi '}{2T_{c}}\right)
-\mu\int\limits_{-E_{F}}^{E_{F}}d{\xi '}
\frac{\Delta(\xi ')}{2\xi '}
th\left(\frac{\xi '}{2T_{c}}\right)
\label{6}
\end{equation}
for $|\xi|<\omega_{c}$,\ while for $\omega_{c}<|\xi|<E_{F}$
\begin{equation}
\Delta(\xi)=-\mu\int\limits_{-E_{F}}^{E_{F}}d{\xi '}
\frac{\Delta(\xi ')}{2\xi '}
th\left(\frac{\xi '}{2T_{c}}\right)
\label{7}
\end{equation}
where we have defined the usual dimensionless constants of pairing and
repulsive interactions:\ $g=N(0)V$, $\mu=N(0)U$.

The general solution of Eqs.(\ref{6}), (\ref{7}) takes the form:
\begin{equation}
\Delta(\xi)=\left\{
\begin{array}{lcl}
\Delta_{c}cos\left(\frac{\pi}{2}\frac{\xi}{\omega_{c}}\right)
+\Delta_{s}sin\left(\frac{\pi}{2}\frac{\xi}{\omega_{c}}\right)
+\Delta &\mbox{ for }&|\xi|<\omega_{c}\\
\Delta &\mbox{ for }&\omega_{c}<|\xi|<E_{F}
\end{array}
\right.
\label{8}
\end{equation}
where $\Delta_{c}$, $\Delta_{s}$, $\Delta$ are determined by the following
algebraic system of equations:
\begin{equation}
\left\{
\begin{array}{lll}
\Delta_{c}=gF_{c}\Delta_{c}+gF\Delta\\
\Delta=-\mu F\Delta_{c}-\mu\ W'\Delta
\end{array}
\right.
\label{9}
\end{equation}

\begin{equation}
\Delta_{s}=gF_{s}\Delta_{s}
\label{10}
\end{equation}
where:
\begin{equation}
\begin{array}{rcl}
F_{c}&=&\int\limits_{0}^{\omega_{c}}d{\xi}cos^{2}\left(\frac{\pi}{2}
\frac{\xi}{\omega_{c}}\right)\frac{1}{\xi}
th\left(\frac{\xi}{2T_{c}}\right)\\
F_{s}&=&\int\limits_{0}^{\omega_{c}}d{\xi}sin^{2}\left(\frac{\pi}{2}
\frac{\xi}{\omega_{c}}\right)\frac{1}{\xi}
th\left(\frac{\xi}{2T_{c}}\right)\\
F&=&\int\limits_{0}^{\omega_{c}}d{\xi}cos\left(\frac{\pi}{2}
\frac{\xi}{\omega_{c}}\right)\frac{1}{\xi}
th\left(\frac{\xi}{2T_{c}}\right)\\
W&=&\int\limits_{0}^{\omega_{c}}d{\xi}\frac{1}{\xi}
th\left(\frac{\xi}{2T_{c}}\right)\\
W'&=&\int\limits_{0}^{E_{F}}d{\xi}\frac{1}{\xi}
th\left(\frac{\xi}{2T_{c}}\right)
\end{array}
\label{FW}
\end{equation}

We can see that the equation (\ref{10}),\ determining $T_{c}$ for the ``odd''
pairing is decoupled from the equations (\ref{9}),\ determining $T_{c}$ for
the ``even'' case.\ The repulsive part of interaction influences only on the
``even'' pairing and from Eq.(\ref{9}) we get the following equation for
$T_{c}$:
\begin{equation}
1=gF_{c}-g\mu\frac{F^{2}}{1+\mu W'} \label{15}
\end{equation}
which may be rewritten as:
\begin{equation}
1=gF_{c}-\mu^{*}W+\mu^{*}g(F_{c}W-F^{2}) \label{15'}
\end{equation}
where we have introduced the Coulomb pseudopotential:\  $$\mu^{*}=\frac{\mu}{1
+\mu (W'-W)}$$ where in the weak-coupling region,\ when
 $T_{c}\ll\omega_{c}$ the value of $W'-W$ reduces to the usual
 $ln\frac{E_{F}}{\omega_{c}}$.

Transition temperature for the ``odd'' state is determined by:
\begin{equation}
1=gF_{s}=g\int\limits_{0}^{\omega_{c}}d{\xi}sin^{2}\left(\frac{\pi}{2}
\frac{\xi}{\omega_{c}}\right)\frac{1}{\xi}th\left(\frac{\xi }{2T_{c}}\right)
\label{16}
\end{equation}

In the Appendix we derive Eqs.(\ref{15}) and (\ref{16}) from the usual
picture of Cooper instability of the normal state.

In Fig.1 we show the results of numerical solution of Eqs.(\ref{15}) and
(\ref{16}) for different values of coupling constants $g$ and $\mu$. For
weak repulsive interaction  ``even'' pairing dominates and the appropriate
transition temperature is higher than the transition temperature for ``odd''
pairing.\ As repulsion grows and for larger values of $g$ the ``odd'' pairing
overcomes the usual one.\  Note that for the model interaction of Eq.
(\ref{V2-cos}) we have the critical value of the pairing constant and the
``odd'' pairing appears only for $g>g_{c}\approx 1.213$.\ Thus we actually
move from the region of weak coupling for which the BCS equations are derived.
In this sense ,\ the results shown in Fig.1 for the region of large coupling
constants are more or less of illustrative nature only.\ In particular,\
this concerns the practically linear growth of $T_{c}$ with the growth of
 $g$,\ which is obvious from Fig.1 and is apparently due to uncontrolled
extrapolation of BCS equations to the region of strong coupling and large $g$.
In fact  here we must use much more elaborate analysis in the spirit of Ref.
\cite{NS},\ where the smooth crossover from ``large'' Cooper pairs of the
weak coupling theory to ``compact'' Bosons of the strong coupling region was
first demonstrated.\ It is known,\ that in this case the growth of $g$ leads
to the saturation of  $T_{c}$,\ which for the large $g$ region is determined
by the well known expression for the temperature of Bose condensation in an
ideal gas of Bosons with practically no dependence on $g$.

The critical value of the coupling constant $g_{c}$ for the ``odd'' pairing is
formally absent if we use the pairing potential of Eq.(\ref{V2-2/3}),\
which is obviously due to its divergence for $|\xi-\xi'|\rightarrow 0$.\
The appropriate dependence of $T_{c}$ on $g=N(0)V$ can be obtained by direct
numerical solution of Eq.(\ref{5}) with potential given in Eq.(\ref{V2-2/3})
and is shown on the inset in Fig.1.\ At the same time it is clear that in this
case also the ``odd'' pairing starts to dominate over ``even'' only in case of
strong enough repulsion.\ Note,\ that the analysis of strong repulsion within
the framework of BCS model is again rather doubtful,\ because it takes into
account only the simplest Fock correction due to electron-electron interaction.
It is clear then that we cannot hope to get consider the limit of
$\mu\rightarrow\infty$ rigorously within this approach.\ However,\ the above
analysis of BCS equation apparently produces the correct qualitative
description of the crossover from traditional ``even'' to the ``odd''pairing.

Consider now the temperature dependence of the gap function for the case of the
``odd'' pairing for the model interaction of Eq.(\ref{V2-cos}).\ According to
Eq.(\ref{8}) for the ``odd'' case we have:
\begin{equation} \Delta(\xi)=\left\{
\begin{array}{lcl}
\Delta_{0}(T)sin\left(\frac{\pi}{2}\frac{\xi}{\omega_{c}}\right)&\mbox{ for
}&|\xi|<\omega_{c}\\ 0&\mbox{ for }&|\xi|>\omega_{c}
\end{array} \right.
\label{17}
\end{equation}
and the temperature dependence of $\Delta_{0}(T)$ is determined by the
following equation which is easily obtained from Eq.(\ref{1}):
\begin{equation}
1=g\int\limits_{0}^{\omega_{c}}d{\xi'}sin^{2}\left(\frac{\pi}{2}
\frac{\xi'}{\omega_{c}}\right)\frac{th\biggl(\frac{\sqrt{\xi'^{2}+
\Delta_{0}^{2}(T)sin^{2}\left(\frac{\pi}{2}\frac{\xi'}{\omega_{c}}\right)}}
{2T}\biggr)}{\sqrt{\xi'^{2}+\Delta_{0}^{2}(T)sin^{2}\left(\frac{\pi}{2}
\frac{\xi'}{\omega_{c}}\right)}}
\label{18}
\end{equation}
Solutions of this equation for the number of values of the pairing constant
$g$ are shown in Fig.2.\ The temperature dependence of $\Delta_{0}(T)$
resemble that of the BCS theory but is not identical to it.\ In particular,\
for the large coupling constants $g\gg g_{c}$ the value of
$2\Delta_{0}(T=0)/T_{c}\approx 5$ with some tendency to become smaller as $g$
diminishes.

The tunneling density of states is easily calculated in a usual way\cite{MA}.\
Using (\ref{17}),\ we get:
\begin{equation}
\frac{N(E)}{N_{0}(0)}=\left\{
\begin{array}{lcl}
\frac{E}{\epsilon+\frac{\pi}{4}\frac{\Delta_{0}^{2}(T)}{\omega_{c}}
sin\left(\pi\frac{\epsilon}{\omega_{c}}
\right)}&\mbox{ for }&|\epsilon|<\omega_{c}\\
1&\mbox{ for }&\omega_{c}<|\epsilon|
\end{array}
\right.
\label{20}
\end{equation}
where $\epsilon$ is determined from the equation
$\epsilon^{2}+\Delta_{0}^{2}(T)sin^{2}\left(\frac{\pi}{2}\frac{\epsilon}
{\omega_{c}}\right)=E^{2}$.
The appropriate dependences for different temperatures are shown in Fig.3.\
Density of states is always gapless,\ ``pseudogap'' is smeared as temperature
grows,\ at the same time the positions of peaks are only weakly dependent on
temperature.\ These results are qualitatively close to those obtained in Ref.
\cite{MA} for the case of interaction given by Eq.(\ref{V2-2/3}) and can be
compared with the known anomalies of the gap-behavior in high-temperature
superconductors.\ If we define the ratio $\frac{2\Delta}{T_{c}}$ using the
peaks
positions of the tunneling density of states we find
$\frac{2\Delta}{T_{c}}\approx 6$.

\section{THE INFLUENCE OF NORMAL IMPURITIES}

The major interest is to study the influence of normal (nonmagnetic) impurities
on the ``odd'' pairing.\ It is well known\cite{Genn,AG1} that such disorder
practically does not influence traditional ``even'' pairing.\ In our case the
equations for normal and anomalous Green's functions have the usual form
\cite{AG1},\ which is valid in case of weak impurity scattering:
\begin{eqnarray}
G(\omega \xi)=-\frac{i\tilde\omega +
\xi}{\tilde\omega^{2}+\xi^{2}+|\tilde\Delta(\xi)|^{2}} \nonumber \\ F(\omega
\xi)=\frac{\tilde\Delta^{\star}(\xi)}{\tilde\omega^{2}+\xi^{2}
+|\tilde\Delta(\xi)|^{2}} \label{21}
\end{eqnarray}
where $\omega=(2n+1)\pi T$,
\begin{eqnarray}
\tilde\omega=\omega-\frac{\gamma}{\pi}\int\limits_{-\infty}^{\infty}d\xi
\frac{\tilde\omega}{\tilde\omega^{2}+\xi^{2}+|\tilde\Delta(\xi)|^{2}}
\nonumber\\
\tilde\Delta(\xi)=\Delta(\xi)+\frac{\gamma}{\pi}\int\limits_{-\infty}^{\infty}
d\xi \frac{\tilde\Delta(\xi)^{\star}}{\tilde\omega^{2}+\xi^{2}
+|\tilde\Delta(\xi)|^{2}}=\Delta(\xi)
\label{22}
\end{eqnarray}
Here $\gamma=\pi cV^{2}_{0}N(0)$---is the scattering rate of electrons due to
point-like impurities with potential $V_{0}$,\ chaotically distributed with
spatial concentration $c$.\ The integral in the second equation in (\ref{22})
is actually zero due to the odd nature of $\Delta(\xi)$,\ so that gap function
renormalization due to impurity scattering is absent.\ This is the main reason
of strong impurity effects in case of the ``odd''-pairing.\ Note that the
similar situation is characteristic for anisotropic,\ e.g.\ $d$-wave pairing
\cite{SS,MBP}.

Gap equation takes now the following form:
\begin{equation}
\Delta(\xi)=-N(0)T\sum_{\omega_{n}}\int\limits_{-\infty}^{\infty}d\xi'
V_{2}(\xi,\xi')
\frac{\Delta^{\star}(\xi')}{\tilde\omega^{2}+\xi'^{2}+|\Delta^{2}(\xi')|^{2}}
\label{23}
\end{equation}
Close to $T_{c}$ this equation can be linearized:
\begin{equation}
\Delta(\xi)=-N(0)T\sum_{\omega_{n}}\int\limits_{-\infty}^{\infty}d\xi'
V_{2}(\xi,\xi')
\frac{\Delta(\xi')}{\tilde\omega^{2}+\xi'^{2}}
\label{24}
\end{equation}
Here $\tilde\omega=\omega+\gamma sign\omega$.

The sum over Matsubara frequencies in (\ref{24}) can be calculated in the usual
way transforming into integral in the complex frequency plane.\ As a result,\
the linearized gap equation can be written in several equivalent forms:
\begin{equation}
\Delta(\xi)=-N(0)\int\limits_{-\infty}^{\infty}d\xi' V_{2}(\xi,\xi')
\int\limits_{-\infty}^{\infty}\frac{d\omega}{\pi}th\left(\frac{\omega}{2T}
\right)ReG^{R}(-\omega\xi')ImG^{R}(\omega\xi')\Delta(\xi')
\label{25}
\end{equation}
where $G^{R}(\omega\xi)=\{\omega -\xi +i\gamma\}^{-1}$---is the retarded
Green's
function of a normal metal with impurities.\ In another way we may write:
\begin{equation}
\Delta(\xi)=-N(0)\int\limits_{-\infty}^{\infty}d\xi' V_{2}(\xi,\xi')
\int\limits_{-\infty}^{\infty}\frac{d\omega}{2\pi}\frac{1}{\xi'}
th\left(\frac{\omega + \xi'}{2T}\right)
\frac{\gamma}{\omega^2+\gamma^{2}}\Delta(\xi')
\label{26}
\end{equation}

Equation similar to (\ref{25}) was recently obtained in Ref.(\cite{MBP}) for
superconductors with anisotropic $d$-wave pairing,\ in the following we use
it in the form of Eq.(\ref{26}).

For the model interaction of Eq.(\ref{V2-cos}) the gap again takes the form
given in Eq.(\ref{17}) and $T_{c}$-equation directly follows from
Eq.(\ref{26}):
\begin{equation}
1=g\int\limits_{0}^{\omega_{c}}\frac{d\xi'}{\xi'}sin^{2}\left(\frac{\pi}{2}
\frac{\xi'}{\omega_{c}}\right)\int\limits_{-\infty}^{\infty}\frac{d\omega}{\pi}
th\left(\frac{\omega+\xi'}{2T_{c}}\right)\frac{\gamma}{\omega^{2}+\gamma^{2}}
\label{27}
\end{equation}
In Fig.4 we show the dependences of $T_{c}$ on $\gamma$ for a number of values
of the pairing coupling constant $g$ which were obtained solving Eq.(\ref{27}).
It is seen that the normal impurity scattering strongly suppresses the ``odd''
pairing.\ Superconductivity is completely destroyed for $\gamma\sim T_{c0}$,\
where $T_{c0}$---is the transition temperature in the absence of scattering
($\gamma\rightarrow 0$),\ defined from Eq.(\ref{16}).\ The destruction of
superconductivity occurs here even faster than in case of introduction of
magnetic impurities into traditional superconductor\cite{AG2}.\ In particular
this is demonstrated by the fast disappearance of superconductivity region on
the ``phase diagram'' of Fig.4 as $g\rightarrow g_{c}$ and also by the absence
of universal dependence of $T_{c}(\gamma)$,\ characteristic for the case of
magnetic impurities.

In case of the model interaction given in Eq.(\ref{V2-2/3}) the $T_{c}$
dependence on $\gamma$ can be obtained by direct numerical solution of the
linearized equation (\ref{26}).\ To calculate the minimal characteristic value,
determining the coupling constant $g$ for a given $T$ we have used the traces
and Kellog's methods\cite{1}.\ Integrals containing $\sim |\xi-\xi'|^{-2/3}$
were calculated by methods which were proposed for singular integrals\cite{2},\
allowing to estimate such integrals with an accuracy of the order of usual
Gaussian quadratures.\ The procedure of calculation of minimal characteristic
values was rather sensitive to the accuracy of calculation of symmetrized
kernels.\ Satisfactory results were obtained using the representation of these
kernels via hypergeometric functions,\ which were calculated by summing the
appropriate generating series with a given accuracy.\ Kellog's method while
being pretty fast in comparison to method of traces has a tendency to
instability in the region of small coupling constants.\ The results obtained
for $T_{c}(\gamma)$ are shown in Fig.5.\ It is obvious that the qualitative
picture obtained from the simplified model interaction above is conserved for
this more ``realistic'' interaction.

The critical scattering $\gamma_{c}$,\ corresponding to the total destruction
of superconductivity ($T_{c}(\gamma\rightarrow\gamma_{c})\rightarrow 0$),\
is determined by the following integral equation:
\begin{equation}
\Delta(\xi)=-N(0)\int\limits_{-\infty}^{\infty}d\xi' V_{2}(\xi,\xi')
\frac{1}{\pi\xi'}arctg\left(\frac{\xi'}{\gamma_{c}}\right)\Delta(\xi')
\label{28}
\end{equation}
which for the interaction given in Eq.(\ref{V2-cos}) reduces to:
\begin{equation}
1=\frac{2}{\pi}g\int\limits_{0}^{\omega_{c}}\frac{d\xi'}{\xi'}sin^{2}\left(
\frac{\pi}{2}\frac{\xi'}{\omega_{c}}\right)arctg\left(\frac{\xi'}
{\gamma_{c}}\right)
\end{equation}
For $g\simeq g_{c}$ from here we can get the dependence like
$\gamma_{c}\sim (g-g_{c})\rightarrow 0$,\ which describes the disappearance of
superconductivity region in Fig.4.\ For $g\gg g_{c}$ (strong coupling) we get
the universal result:\ $\gamma_{c}/T_{c0}=4/\pi\approx 1.273$.\ In fact this
result as well as the dependence of $T_{c}(\gamma)$ for $g\gg g_{c}$ are
independent from the choice of model potential $V_{2}(\xi,\xi')$.\
Concerning the universality of the ratio $\gamma_{c}/T_{c0}$,\ it is easy to
see that it follows from the equivalent form of Eq.(\ref{16}) for $T_{c0}$
and Eq.(\ref{28}) for $\gamma_{c}$ in case of
$T_{c0}\gg\omega_{c}$ and $\gamma_{c}\gg\omega_{c}$ (i.e. for $g\gg g_{c}$):
\begin{equation}
\Delta(\xi)=-\frac{N(0)}{4T_{c0}}\int\limits_{-\infty}^{\infty}d\xi'
V_{2}(\xi,\xi')\Delta(\xi') \label{30}
\end{equation}
\begin{equation}
\Delta(\xi)=-\frac{N(0)}{\pi\gamma_{c}}\int\limits_{-\infty}^{\infty}d\xi'
V_{2}(\xi,\xi')\Delta(\xi') \label{31}
\end{equation}
The equivalence of these equations allows us to equate the coefficients of
integrals which immediately gives the above result for the ratio of
$\gamma_{c}/T_{c0}$. Analogously it is easily seen that Eq.(\ref{26})
for $T_{c}(\gamma)\gg\omega_{c}$ reduces to:
\begin{equation}
\Delta(\xi)=-\frac{N(0)}{4T_{c}(\gamma)}f\left(\frac{\gamma}{T_{c}(\gamma)}
\right)\int\limits_{-\infty}^{\infty}d\xi' V_{2}(\xi,\xi')\Delta(\xi')
\label{32}
\end{equation}
where $$f(x)=\int\limits_{-\infty}^{\infty}\frac{d\omega}{\pi}
\frac{1}{ch^{2}(\omega/2)}\frac{x}{x^{2}+\omega^{2}}$$
Accordingly,\ comparing Eqs. (ref{32}) and (\ref{30}) we see that in the limit
of strong coupling the dependence of $T_{c}$ on $\gamma$ is determined by the
following universal equation:
\begin{equation}
\frac{T_{c}(\gamma)}{T_{c0}f\left(\frac{\gamma}{T_{c}(\gamma)}\right)}=1
\label{33}
\end{equation}
We must stress however,\ that these ``strong coupling'' results are of rather
formal nature and should be seriously modified in the spirit of Ref.\cite{NS}.

Our results for $\gamma_{c}$ for the case of the model interaction
of Eq.(\ref{V2-cos}),\ and those obtained by numerical solution of
Eq.(\ref{28})
with ``realistic'' interaction of Eq.(\ref{V2-2/3}) are shown at the insets
in Fig.4 and Fig.5.

\section{CONCLUSION}

In conclusion let us formulate the main results.\ We have presented the simple
model of the pairing interaction which allows the complete analysis of
integral equations of the BCS theory both for the usual ``even''
(over $k-k_{F}$) and for the exotic ``odd'' pairing.\ It is shown that the
``odd'' pairing becomes preferable for sufficiently strong electron repulsion
and,\ in general,\ for sufficiently strong pairing interaction.\ The last
problem (``strong coupling'') deserves further studies of the crossover from
Cooper pairs to compact Bosons.\ The ``odd'' pairing leads to the picture of
gapless superconductivity and to other deviations from the traditional BCS
theory such as the unusual evolution of the pseudogap in the density of states,
the large ratio of $2\Delta_{0}/T_{c}$ etc.,\ which are attractive from the
point of view of the theory of high-temperature superconductors.

At the same time normal impurities (disorder) strongly suppress the ``odd''
pairing.\ This effect is even stronger,\ than in the case of magnetic
impurities
in usual superconductors.\ Even in the ``strong coupling'' limit
superconductivity is completely destroyed for $\gamma\sim T_{c0}$,\ while for
the smaller pairing constants superconductivity region on the phase diagram
rapidly shrinks.

It is rather well known that high-temperature superconductors are unstable
with respect to structural disordering\cite{Aleks}.\ However,\ if we exclude
special cases like the impurities of Zn,\ superconductivity is suppressed in
metallic oxides rather close to the disorder-induced metal-insulator
transition,\ which is apparently due to Anderson localization\cite{Aleks}.\
Anderson transition takes place for $\gamma\sim E_{F}\gg T_{c0}$,\ so that the
``odd'' pairing is destroyed well before.\ Apparently this makes the ``odd''
pairing rather unlikely explanation of high-temperature superconductivity in
oxides.\ However,\ we can not exclude the possibility that some effect in these
systems can be explained by the fast suppression of the ``odd'' component of
superconducting order parameter by disordering,\ while the ``even'' part which
is stable to disordering remains.\ This problem deserves further studies.

\vskip 0.5cm
This work was supported by the Scientific Council on High-Temperature
Superconductivity under the Project $N^{o}$ 90135 of the State Research
Program on Superconductivity.\ It was also partly supported by the Soros
Foundation grant awarded by the American Physical Society.

\newpage
\begin{center}
{\bf APPENDIX }
\end{center}

\vskip 1.0cm
It is useful to derive equations determining $T_{c}$ both for ``even'' and
``odd'' pairings from the picture of normal state instability,\ i.e. as the
points of the relevant Cooper instabilities.\ Consider the two-particle
Green's function in Cooper channel which is shown diagrammatically in Fig.6.\
It is convenient to analyze instability of the normal state associating it
with the divergence of this Green's function,\ summed over Matsubara
frequencies:
\begin{equation}
\Phi_{pp'}(\Omega q)=-T\sum_{\omega\omega'}\Phi_{pp'}(\omega\omega'\Omega q)
\end{equation}
for $q=\Omega=0$.\ Consider again electron-electron interaction
$V(\xi,\xi')$ consisting of two parts defined in Eqs.(\ref{V1}) and
(\ref{V2-cos}).\ Due to isotropy of the system $\Phi_{pp'}(00)$ may be
represented as $\Phi(\xi,\xi')$,\ which is defined by the equation:
\begin{equation}
\Phi(\xi,\xi')=Z(\xi)\delta_{\xi\xi'}+Z(\xi)N(0)\int\limits_{-\infty}^{\infty}
d\xi'V(\xi-\xi')\Phi(\xi,\xi')
\end{equation}
where
\begin{equation} Z(\xi)=-T\sum_{\omega}G(\omega\xi)G(-\omega\xi)=
-\frac{1}{2\xi}th\left(\frac{\xi}{2T}\right)
\end{equation}
Taking into account Eqs.(\ref{V1}) and (\ref{V2-cos}) we have:
\begin{eqnarray}
\Phi(\xi,\xi')=Z(\xi)\delta_{\xi\xi'}+Z(\xi)\left\{\mu\int
\limits_{-E_{F}}^{E_{F}}d\xi'\Phi(\xi,\xi')-\right.\nonumber\\
-gcos\left(\frac{\pi}{2}\frac{\xi}{\omega_{c}}\right)
\int\limits_{-\omega_{c}}^{\omega_{c}}
d\xi'cos\left(\frac{\pi}{2}\frac{\xi'}{\omega_{c}}\right)\Phi(\xi,\xi')-
\nonumber\\
\left.-gsin\left(\frac{\pi}{2}\frac{\xi}{\omega_{c}}\right)
\int\limits_{-\omega_{c}}^{\omega_{c}}
d\xi'sin\left(\frac{\pi}{2}\frac{\xi'}{\omega_{c}}\right)\Phi(\xi,\xi')\right\}
\label{A4}
\end{eqnarray}
for $|\xi|,|\xi'|<\omega_{c}$ and,\ accordingly:
\begin{equation}
\Phi(\xi,\xi')=Z(\xi)\delta_{\xi\xi'}+Z(\xi)\mu\int\limits_{-E_{F}}^{E_{F}}
d\xi'\Phi(\xi,\xi')
\label{A5}
\end{equation}
for $|\xi|$ or $|\xi'|>\omega_{c}$, $|\xi|,|\xi'|<E_{F}$.\\ Let us define the
functions:
\begin{equation}
\begin{array}{rcl}
f_{c}(\xi)&=&\int\limits_{-\omega_{c}}^{\omega_{c}}
d\xi'cos\left(\frac{\pi}{2}\frac{\xi'}{\omega_{c}}\right)\Phi(\xi,\xi')\\
f_{s}(\xi)&=&\int\limits_{-\omega_{c}}^{\omega_{c}}
d\xi'sin\left(\frac{\pi}{2}\frac{\xi'}{\omega_{c}}\right)\Phi(\xi,\xi')\\
f(\xi)&=&\int\limits_{-E_{F}}^{E_{F}}d\xi'\Phi(\xi,\xi')
\end{array}
\end{equation}
for which,\ using Eqs.(\ref{A4}),\ (\ref{A5}),\ we obtain the following system
of equations:
\begin{equation}
\left\{
\begin{array}{rcl} f(\xi)&=&Z(\xi)-\mu W'f(\xi)+gFf_{c}(\xi)\\
f_{c}(\xi)&=&Z(\xi)cos\left(\frac{\pi}{2}\frac{\xi}{\omega_{c}}\right)-\mu
Ff(\xi)+gF_{c}f_{c}(\xi)\\
f_{s}(\xi)&=&Z(\xi)sin\left(\frac{\pi}{2}\frac{\xi}{\omega_{c}}\right)
+gF_{s}f_{s}(\xi)
\end{array}
\right.  \label{A8}
\end{equation}
where we have used notations introduced in (\ref{FW}).\\ Now we can see that
the ``odd'' and ``even'' equations has decoupled.\ The ``odd'' pairing is
connected with the divergence of the function $f_{s}(\xi)$,\ and the
corresponding instability condition has the form of $1=gF_{s}$,\ which
coincides with Eq.(\ref{16}).\ The first two equations in (\ref{A8})
determine the instability towards ``even'' pairing.\ It is easy to see that
$$f_{c}(\xi)=Z(\xi)\left\{cos\left(\frac{\pi}{2}\frac{\xi}{\omega_{c}}\right)
-\frac{\mu F}{1+\mu W'}\right\}\left\{1-gF_{c}+\frac{g\mu F^{2}}{1+\mu W'}
\right\}^{-1}$$ and instability condition takes the following form:
\begin{equation}
1=gF_{c}-g\mu\frac{F^{2}}{1+\mu W'}
\end{equation}
which coincides with Eq.(\ref{15}).

\newpage
\begin{center}
{\bf Figure Captions:}
\end{center}

\vskip 0.5cm
Fig.1.\ Dependence of $T_{c0}$ on the pairing coupling constant $g=N(0)V$ in
case of the ``even'' (lines) and ``odd'' (dashed lines) pairings.

1---$\mu=0$;\ 2---$\mu=1$;\ 3---$\mu=10$.\ In numerical procedures it
was assumed that $E_{F}/\omega_{c}=50$.

At the inset---analogous dependence for ``realistic'' interaction of Eq.(3).

\vskip 0.5cm
Fig.2.\ Temperature dependence of the gap $\Delta_{0}(T)$ in case of the
``odd'' pairing for the number of values of pairing coupling constant.

1---$g=1.22$;\ 2---$1.5$;\ 3---$2.0$;\ 4---$3.0$;\ 5---$4.0$;\ 6---$5.0$.

\vskip 0.5cm
Fig.3.\ Density of states in the model of the ``odd'' pairing for a number of
characteristic temperatures.

1---$T/T_{c}=0$;\ 2---$0.6$;\ 3---$0.8$;\ 4---$0.9$;\ 5---$0.99$.

Pairing coupling constant was assumed to be $g=3$.

\vskip 0.5cm
Fig.4.\ Dependence of $T_{c}$ for the ``odd'' pairing on the impurity
scattering rate $\gamma$ for different values of the coupling constant $g$:

1---$g=1.22$;\ 2---$1.24$;\ 3---$1.30$;\ 4---$1.50$;\ 5---$2.0$;\
6---$5.0$;\ 7---$10.0$.

At the inset---dependence of the critical scattering rate on the pairing
coupling constant.

\vskip 0.5cm
Fig.5.\ Dependence of $T_{c}$ for the ``odd'' pairing on the impurity
scattering rate $\gamma$ for different values of the pairing coupling
constant $g$ in the model with ``realistic'' interaction of Eq.(3):

1---$g=0.17$;\ 2---$0.25$;\ 3---$0.72$;\ 4---$1.15$;\ 5---$6.41$.

At the inset---dependence of the critical scattering rate on the pairing
coupling constant

\vskip 0.5cm
Fig.6.\ Two-particle Green's function in Cooper channel.
\end{document}